# Mixed structural face-centered cubic and body-centered cubic orders in near stoichiometric Fe2MnGa alloys

Y. V. Kudryavtsev, A. E. Perekos, N. V. Uvarov, M. R. Kolchiba, K. Synoradzki, and J. Dubowik

### Articles you may be interested in

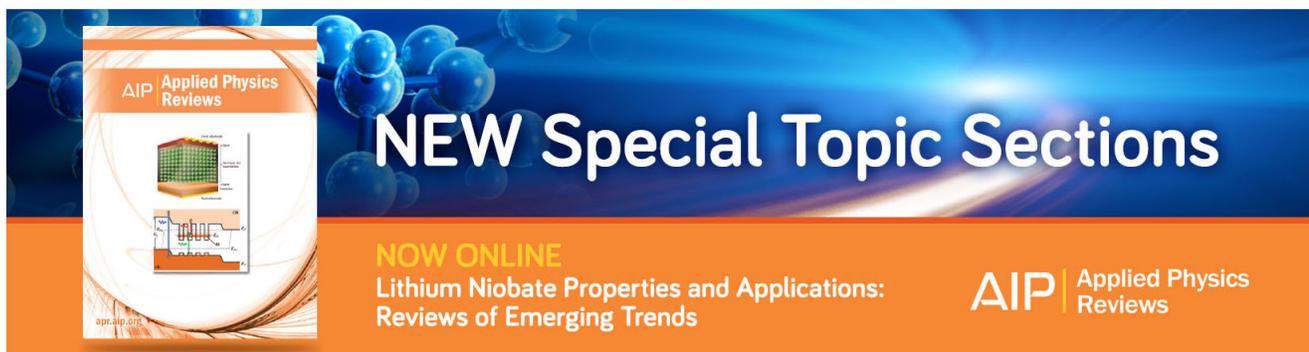

# Mixed structural face-centered cubic and body-centered cubic orders in near stoichiometric Fe$_2$MnGa alloys

Y. V. Kudryavtsev,[1,2] A. E. Perekos,[1] N. V. Uvarov,[1] M. R. Kolchiba,[1] K. Synoradzki,[3] and J. Dubowik[3]
[1]*Institute of Metal Physics, NAS of the Ukraine, Vernadsky 36, 03680 Kiev-142, Ukraine*
[2]*Donetsk Physical and Technical Institute, NAS of the Ukraine, Nauki 46, 03028 Kiev, Ukraine*
[3]*Institute of Molecular Physics, Polish Academy of Sciences, Smoluchowskiego 17, 60-179 Poznań, Poland*

Magnetic and transport properties of near stoichiometric metastable Fe$_x$Mn$_y$Ga$_z$ alloys ($46 \leq x \leq 52$, $17 \leq y \leq 25$, $26 \leq z \leq 30$) with face-centered cubic (FCC), body-centered cubic (BCC), and two-phase (FCC + BCC) structures are investigated. The experimental results are analyzed in terms of first-principles calculations of stoichiometric Fe$_2$MnGa alloy with the L2$_1$, L1$_2$, and the tetragonally distorted L2$_1$ structural orderings. It is shown that the pure BCC and FCC phases have distinct magnetic and transport properties. Two-phase Fe$_2$MnGa alloys have magnetic and transport properties typical of the mixed BCC and FCC phases. Among the investigated alloys, Fe$_{46}$Mn$_{24}$Ga$_{30}$ has a martensitic transformation accompanied with significant changes of its magnetic and transport properties. *Published by AIP Publishing.* [http://dx.doi.org/10.1063/1.4952392]

## INTRODUCTION

Significant interest in stoichiometric and off-stoichiometric Fe$_2$MnGa Heusler alloys (HAs) has arisen due to the many effects discovered to date: a martensitic transformation,[1–5] a metamagnetic transformation from an antiferromagnetic (AFM) to a ferromagnetic (FM) phase,[6–9] and a large exchange bias.[6–8]

Unlike Ni-Mn-based HA,[10] the Fe$_2$MnGa alloys can have a more complex magnetic behavior due to the multiple structures in which they may crystallize. For single phase $\gamma$-Fe$_2$MnGa alloys (i.e., alloys with a face-centered cubic (FCC) structure), a FM to AFM transformation has been observed at 220–250 K.[6–9,11] However, as has been found in Fe$_{50}$Mn$_{24}$Ga$_{26}$,[9] even a small deviation from stoichiometric composition results in a mixed body-centered cubic (BCC) and FCC structure, with their respective magnetic transformation from a FM to a paramagnetic (PM) state at 190 K and an AFM to FM transformation at 240 K. In this way, the nearly stoichiometric Fe$_2$MnGa alloys may crystallize either in the BCC or FCC types of structure. Okumura *et al.*[12] have shown that a small change in the melt-spinning processing parameter results in the formation of Fe$_2$MnGa alloys with BCC or FCC crystal structures from the same precursor ingots. It has been experimentally shown by Yin *et al.*[13] that among the Fe-based HAs, Fe$_2$MnGa has the minimal negative enthalpy of formation. The stability of the FCC or BCC phases in Fe$_2$MnGa alloys has been estimated in terms of a valence-electron-to-atom ($e/a$) ratio that determines the structural transformation by affecting the instability of a phonon mode.[14,15]

On the other hand, a distinct martensitic transformation has also been found in slightly off-stoichiometric Fe$_2$MnGa alloys. Zhu *et al.*[1] have observed martensitic transformation in Fe$_{50.0}$Mn$_{22.5}$Ga$_{27.5}$ alloy from a parent austenite PM BCC phase (with the lattice parameter $a = 0.5856$ nm) to a martensite FM body-centered tetragonal (BCT) structure ($a = b = 0.5328$ nm and $c = 0.7113$ nm). Several publications have reported experimental studies of martensitic transformation in slightly off-stoichiometric Fe$_2$MnGa alloys having a BCC parent austenite phase.[2–5] The martensite phase has an L1$_0$ (or D0$_{22}$) structure with the lattice parameters $a = 0.381$ nm and $c = 0.353$ nm[2] or a tetragonal structure with $a = b = 0.537$ nm, $c = 0.708$ nm.[3] It has been also shown that the martensitic transformation in the Fe$_2$MnGa alloys is accompanied with significant changes in their mechanical, magnetic, and transport properties.[1,2,5]

Consequently, one may infer that Fe$_2$MnGa alloys show an unusually rich metastable behavior for a relatively small deviation from the stoichiometric composition: they may exhibit a BCC structure (defined as a disordered L2$_1$ structure), a FCC structure (*i.e.*, a disordered L1$_2$ structure), or they may exhibit a martensitic transformation from a BCC parent phase to a tetragonal structure defined as a tetragonally distorted BCC structure for a certain range of compositions. In this paper, we report the results of magnetic and transport measurements of Fe$_2$MnGa alloys with distinct structural orderings and make a comprehensive analysis on how various types of structural orders affect their magnetic and transport properties.

## EXPERIMENTAL DETAILS

Several bulk polycrystalline Fe$_2$MnGa alloys near the stoichiometry 2:1:1 were prepared by melting together pieces of Fe, Mn, and Ga of 99.99% purity in an arc furnace with a water cooled Cu hearth under a 1.3 bar Ar atmosphere. The Ar gas in the furnace before melting was additionally purified by multiple remelting of a Ti$_{50}$Zr$_{50}$ alloy getter. To promote the volume homogeneity, the ingots were remelted five times. After ingot melting, the weight loss was negligible. The actual alloy compositions were evaluated by using energy dispersive x-ray spectroscopy (see Table I).

TABLE I. Composition of $Fe_2MnGa$ alloys, amounts of FCC and BCC phases, valence electron concentration, resistivity of alloys at $T=300$ K, and saturation magnetization measured at $T=4.2$ K.

| Sample No. | Alloy composition (at. %) | FCC phase amount (vol. %) | BCC phase amount (vol. %) | Valence electron concentration (electron/atom) | $\rho(300\,K)$ $\times 10^{-6}$ ($\Omega$ cm) | Saturation magnetization (emu/g) |
|---|---|---|---|---|---|---|
| 1 | $Fe_{49}Mn_{25}Ga_{26}$ | 100 | 0 | 6.45 | 290 | 120 |
| 2 | $Fe_{50}Mn_{24}Ga_{26}$ | 92 | 8 | 6.46 | 257 | 109 |
| 3 | $Fe_{50}Mn_{23}Ga_{27}$ | 71 | 29 | 6.42 | 152 | 46 |
| 4 | $Fe_{50}Mn_{22}Ga_{28}$ | 39 | 61 | 6.38 | 242 | 61 |
| 5 | $Fe_{46}Mn_{24}Ga_{30}$ | 19 | 81 | 6.27 | 550 | 78 |
| 6 | $Fe_{52}Mn_{18}Ga_{30}$ | 0 | 100 | 6.32 | 356 | 53 |

The structural characterization of the sample was carried out at room temperature (RT) employing x-ray diffraction (XRD) in $\theta - 2\theta$ geometry with Cu-$K_\alpha$ radiation ($\lambda = 0.15406$ nm). An SC 404 F1 Pegasus differential scanning calorimeter (DSC) was used to determine the phase transformation temperatures. The magnetic properties of the bulk $Fe_2MnGa$ alloy samples were investigated over the temperature range $80 \leq T \leq 825$ K by measuring the DC-magnetic susceptibility in a weak magnetic field of 5 Oe and by measuring the alloy magnetization over the temperature range $2 \leq T \leq 400$ K and a range of magnetic fields $0 \leq H \leq 50$ kOe by using the PPMS-P7000 system. The transport properties of the $Fe_2MnGa$ alloy samples were measured by using the four-probe technique over the range of temperatures $80 \leq T \leq 600$ K. The samples for the transport measurements, about $1 \times 1 \times 15$ mm$^3$ in size, were cut from an ingot by using the spark-erosion technique.

Electronic structure calculations (density of states $N(E)$ (DOS), total electron energy, and magnetic properties) of the stoichiometric $Fe_2MnGa$ HA for ordered $L2_1$- (225 space group) and $L1_2$- (123 space group) types of structure (see Fig. 1) as well as for FM and ferrimagnetic (FI) types of magnetic order were calculated by using the WIEN2k code[16] utilizing a full-potential linearized-augmented-plane-wave method.[17] For the exchange-correlation functional, the generalized-gradient-approximation version of Perdew *et al.*[18] was used. Self-consistency was obtained using 816 k-points in the irreducible Brillouin zone. The lattice parameters of the $L2_1$ and $L1_2$ structures obtained as the result of volume optimization were used for the calculations. Additionally, the electronic structure of a tetragonally distorted $L2_1$-phase of $Fe_2MnGa$ alloy was also calculated. The experimental lattice constants for the tetragonal phase[3] were used for the calculations.

## RESULTS AND DISCUSSION

### Electronic structure

Our previous results have shown that the $L2_1$ crystalline structure in $Fe_2MnGa$ alloy is stable only for FI type of magnetic ordering.[19] We have found no solution for the FM ordering in the $L2_1$ structure. On the other hand, for the $L1_2$ crystalline structure of a $Fe_2MnGa$ alloy, both the FM and FI types of magnetic order have been shown to be equally probable.

The calculated DOS of a stoichiometric $Fe_2MnGa$ alloy with tetragonal, $L2_1$-, and $L1_2$ types of atomic order shown in Fig. 2 shows that symmetry significantly affects the electronic structure of the alloy. In particular, strong differences are observed near the Fermi level $E_F$. It should also be noted that the spin-up and spin-down states for all types of structure are strongly polarized, since the separate peaks of the up- and down-states neither coincide with energy nor with intensity (see Fig. 2). The main contribution to the DOS is made by the Fe and Mn atoms, whose states are strongly hybridized, as the most intense DOS peaks are formed by the Fe and Mn states coinciding in the energy scale. The Ga contribution to the total DOS is small, i.e., the Ga atoms basically form ionic bonds with the surrounding atoms.

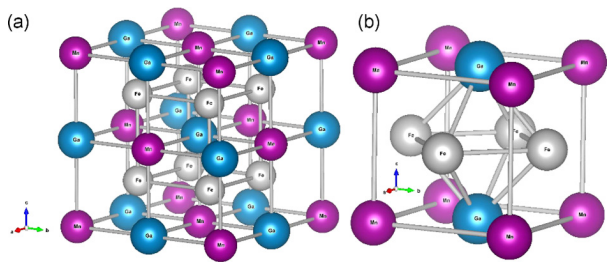

FIG. 1. $L2_1$ (a) and $L1_2$ (b) structures of $Fe_2MnGa$ alloy. The atoms are coded as follows. Fe: gray, Mn: magenta, Ga: blue.

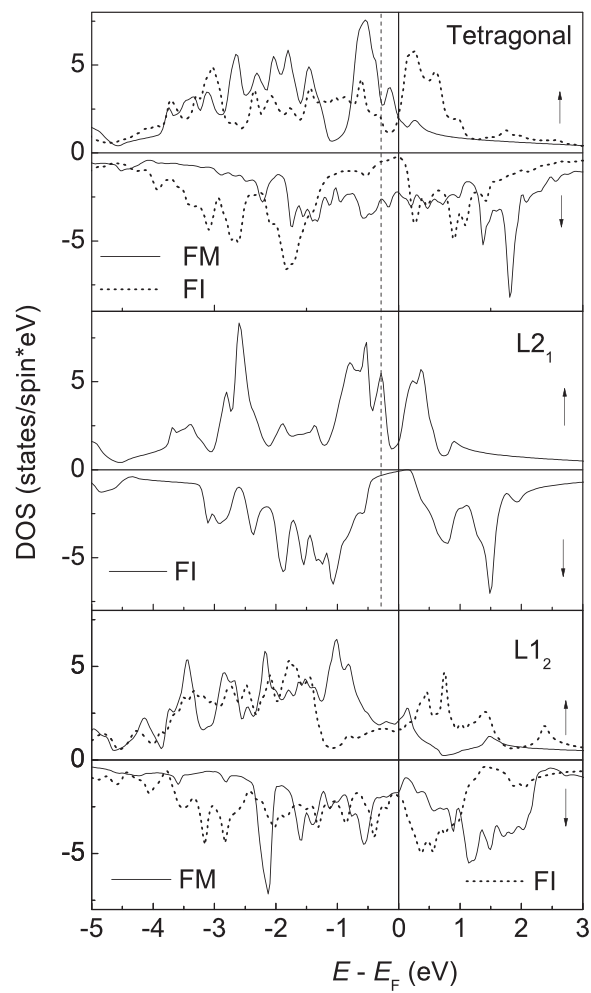

FIG. 2. Density of states of stoichiometric $Fe_2MnGa$ alloy calculated for different types of structural and magnetic order.

TABLE II. Calculated total and partial magnetic moments per formula unit, DOS at $E_F$, and total electron energies of Fe$_2$MnGa alloys with different types of atomic and magnetic orders. Calculated lattice constants are shown together with experimental results.

| Structure | Magnetic order | $\mu_{tot}$ $\mu_B$ | $\mu_{Fe}$ $\mu_B$ | $\mu_{Mn}$ $\mu_B$ | $\mu_{Ga}$ $\mu_B$ | $N(E_F)$ (states/eV spin) | $E_{tot}$ (eV) | $a_{calc}$ nm | $a_{exp}$ nm |
|---|---|---|---|---|---|---|---|---|---|
| Tetragonal | FM | 6.355 | 2.231 | 2.101 | −0.112 | 5.10 | −11296.7456 | | $a=b=0.5368, c=0.7081$ |
| Tetragonal | FI | 0.201 | −1.620 | 3.070 | 0.020 | 2.43 | −11296.7621 | | $a=b=0.5368, c=0.7081$ |
| L2$_1$ | FI | 2.010 | −0.450 | 2.870 | 0.000 | 1.65 | −11296.7502 | 0.5700 | 0.5871 |
| L1$_2$ | FM | 6.130 | 1.980 | 2.290 | −0.090 | 3.84 | −11296.7666 | 0.3644 | 0.3701 |
| L1$_2$ | FI | 0.480 | −1.760 | 2.950 | 0.002 | 3.56 | −11296.7424 | 0.3666 | |

The phase stability of Ni-, Co-, and Fe-based HA has been studied by Entel and Zayak.[14,15] According to them, unstable HA has an intense $N(E)$ peak at $E_F$, which is responsible for the nesting topology of the Fermi surface and the covalent bonding features. This affects the optical vibrational TA$_2$ modes of Ni-based system (unlike those based on Co or Fe).[14,15] From this point of view, the stoichometric Fe$_2$MnGa alloy with L2$_1$ structure and FI magnetic order is stable.[14,15]

Considering the obtained values of $N(E_F)$ and $E_{tot}$ for the different types of atomic and magnetic orders in stoichiometric Fe$_2$MnGa alloy (see Table II), it is hard to deduce its phase stability. Among the calculated structures, the FI ordered L2$_1$ phase of Fe$_2$MnGa alloy has the lowest $N(E_F)$, but its $E_{tot}$ is in the middle. The FM L1$_2$ phase of Fe$_2$MnGa has the lowest $E_{tot}$ but its $N(E_F)$ is nearly twice as big as that for the FI L2$_1$ phase. For these possible types of atomic and magnetic orders of stoichiometric perfectly ordered Fe$_2$MnGa alloy, $\Delta E_{tot} = 0.0164$ eV (or 190 K). For the tetragonal phase of Fe$_2$MnGa alloy, the FI magnetic order is more preferable than the FM.

As can be seen in Table II, the main contribution to the total magnetic moment $\mu_{tot}$ for the cubic phases of Fe$_2$MnGa is by moments localized at the Mn. For the tetragonal phase, the magnetic moments localized at Fe and Mn sites give nearly the same contribution to $\mu_{tot}$. The moment localized at Ga is negligible.

### Structure

Figure 3 shows the XRD experimental patterns of our samples (open circles). The solid (red) lines through the circles are the results of a Ritveld refinement with the differences between the data and refinement shown in the bottom panels for each sample. According to the XRD data for the Fe$_2$MnGa alloys, the three furthest samples 1 (2) and 6 have the disordered L1$_2$ (i.e., FCC) and disordered L2$_1$ (i.e., BCC or A2) types of structures, respectively. The fundamental diffraction peaks for these types of structure are shown by two stick diagrams at the top and bottom panels. In a closer examination of the refinements for samples 1 and 2, it is hard to see whether the pure FCC structure obtains for sample 1 or 2. It seems that sample 1 has the less intensive peak (220) from the BCC phase. However, sample 2 has no (400) peak of L2$_1$, which is seen for sample 1. Therefore, it is possible that sample 2 is more homogeneous. Other samples contain both FCC and BCC phases in different ratios (see Fig. 3 and Table I). The amounts of the FCC and BCC phases in the alloys have been evaluated by the ratios of the most intense (111) [for the FCC phase] and (220) [for the BCC phase] diffraction peaks. The samples in Table I and in most of the figures were arranged by the content of the FCC and BCC phases, with some uncertainty concerning samples 1 and 2. It is characteristic that the XRD pattern of sample 5 has significantly broader reflections than that of the other samples. As will be discussed in the Discussion section, such an XRD pattern may be related to a tetragonally distorted BCC structure of a martensitic phase. The experimental values of the lattice constants for the FCC and BCC phases in $Fe_2MnGa$ alloys are close to those found from the first-principles calculations (see Table II).

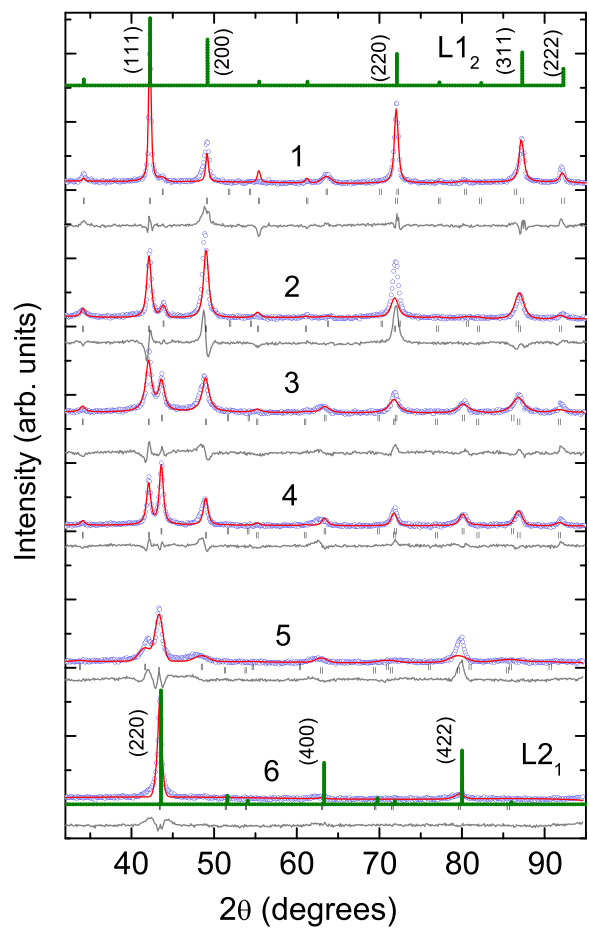

FIG. 3. Experimental (symbols) and simulated (lines) XRD spectra of Fe$_2$MnGa alloys. Numbers indicate Fe$_2$MnGa alloy samples. Stick diagrams (green) show the main XRD lines of the L2$_1$ and L1$_2$ structures.

It is seen that the close probabilities of formation of FI $L2_1$- and FM $L1_2$ phases result in a structural instability of the $Fe_2MnGa$ alloys: An insignificant deviation in the alloy composition from the stoichiometric one results in a change in its phase content. It may be assumed that an increase in Ga content in alloys leads to an increasing probability of the formation of the BCC phase (see Table I).

Figure 4 shows the DSC results of the $Fe_2MnGa$ samples with the regions of interest (A, B, and C) depicted by yellow areas. A large number of not well specified endo- and exo-thermic transformations in the $Fe_2MnGa$ alloys in a temperature range of 400–1100 K can be interpreted as an apparent sign of structural instability of the $Fe_2MnGa$ alloys. In the temperature region "A," one exo- and one endothermic peak are related to the onset of the FCC phase in agreement with Ref. 12. As is shown by the blue dashed lines through 1 and 2, the endothermic peak definitely related to a FCC phase disappears. In contrast, an endothermic peak at about 850 K intensifies and serves as a sign of the onset of a BCC phase. At lower temperatures (region "B"), weak endothermic peaks at 700–800 K are related to a FM/PM transformation. As will be discussed in Fig. 5, the endothermic peaks are the most pronounced for samples 1 and 2 and disappear completely for samples 5 and 6. Eventually, at the lowest temperature range 400–600 K (region "C"), there are some signs (see the discussion of Fig. 5) of a transition attributed to an AFM/FM transformation in the FCC phase. As is depicted by dashed red lines, samples 1 and 2 can be regarded as mostly constituted by the FCC phase with a small admixture of the BCC phase, while samples 5 and 6 are mostly made up of the BCC phase. We conclude that the DSC results are in agreement with the XRD data.

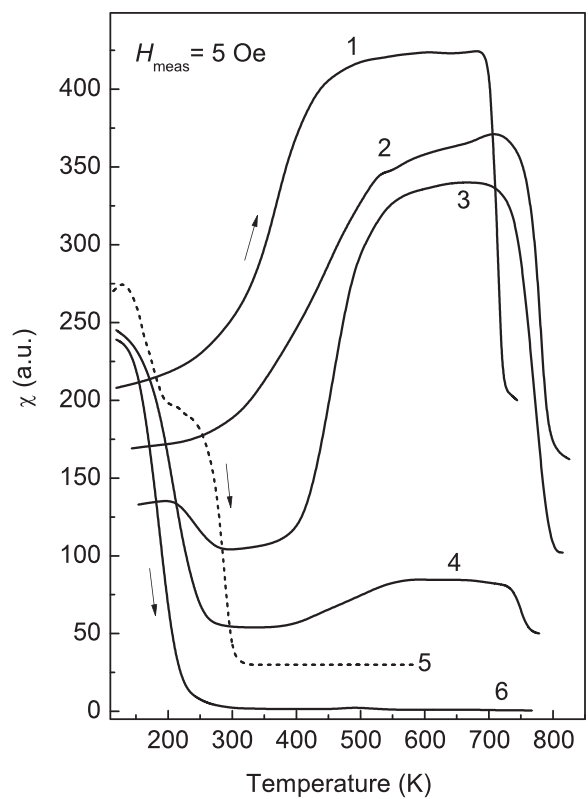

FIG. 5. Temperature dependencies of DC magnetic susceptibility of $Fe_2MnGa$ alloys obtained on warming. The $\chi(T)$ curves of $Fe_2MnGa$ alloys are shifted. Numbers indicate the samples.

## Magnetic properties

As shown in Fig. 5, the FCC and BCC phases in the $Fe_2MnGa$ alloys have distinct magnetic properties, which may be determined from the temperature dependencies of the magnetic susceptibility $\chi(T)$ on warming. While the BCC phase (sample 6) is ferromagnetic below its Curie temperature $T_C \approx 200-220$ K, the FCC phase (samples 1 and 2) is ferromagnetic below $T_C \approx 730-750$ K but its susceptibility decreases rapidly below $T \approx 400$ K. The apparently mixed-phase $Fe_2MnGa$ alloys (samples 3 and 4, for example) have $\chi(T)$ being a combination of the temperature dependencies of susceptibilities of the pure BCC and FCC phases. Similar results have already been obtained (Ref. 12), however, with distinct $T_C$ of 400 K for the FCC phase. In contrast, $\chi(T)$ of sample 5 shows an additional drop on warming at about 260 K.

At higher fields of 50–500 Oe, the temperature dependencies of the magnetization shown in Fig. 6 are more complex. In particular, for samples 4 and 6, a long "tail" is seen at $T > 250$ K both on cooling and warming, while the magnetization of sample 5 reveals a clear hysteretic behavior on cooling and warming.

In contrast, for a $Fe_2MnGa$ alloy containing mainly the FCC phase (samples 1 and 2), the magnetization increases with temperature. Sample 3 contains $\approx 30\%$ of the BCC phase, and its $M(T)$ dependence exhibits the features typical for a mixture of the BCC and FCC phases. It should be emphasized here that $M(T)$ for field cooling and field warming plots for the aforementioned samples do not show any hysteretic behavior.

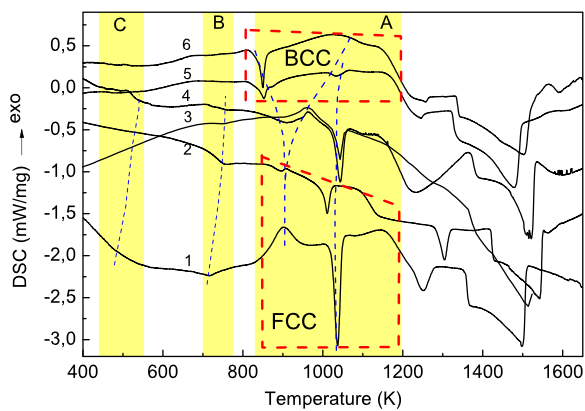

FIG. 4. Differential thermal analysis of bulk $Fe_2MnGa$ alloys. Numbers indicate the alloy samples. Yellow areas depict regions of interest.

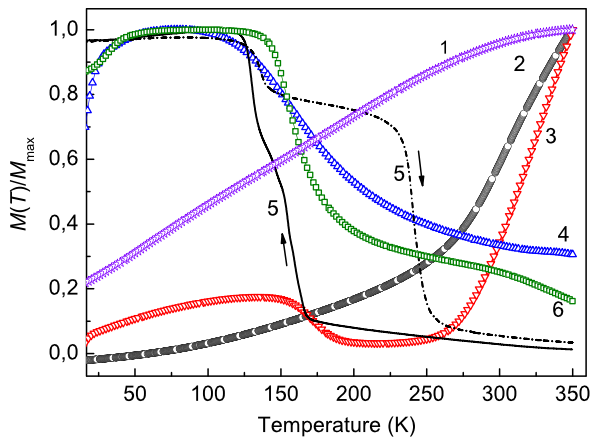

FIG. 6. Temperature dependencies of normalized magnetization of bulk Fe$_2$MnGa alloys taken at low ($H = 50 \div 500$ Oe) magnetic fields. $M(T)/M_{max}$ plots for all the samples (except sample 5) practically coincide for the cooling and warming scans, only the warming scans are presented. Numbers near the $M(T)/M_{max}$ curves indicate alloy samples.

These unusual temperature dependencies of DC magnetic susceptibility (Fig. 5) and magnetization of Fe$_2$MnGa alloys containing the FCC-phase taken at relatively weak measuring magnetic fields (Fig. 6) have been observed earlier and explained as resulting from a metamagnetic transformation from low-temperature AFM to a high-temperature FM state.[6–9] At the same time, Passamani *et al.*[11] have explained the temperature dependence of the magnetization of the FCC Fe$_2$MnGa alloys by different temperature dependencies of magnetization for antiparallel aligned magnetic moments localized at the Fe and Mn sublattices.

A hysteretic behavior of the temperature dependence of the magnetization of sample 5 needs some explanation. In agreement with earlier observations, such an anomaly in $M(T)$ can be definitely ascribed to a martensitic transformation.[1,5] The martensitic transformation is confirmed by independent differential calorimetric measurements. As is seen in Fig. 7, DSC measurements for sample 5 demonstrate clear exothermic and endothermic peaks on cooling and warming, in agreement with the hysteretic behavior of magnetization shown in Fig. 6 for sample 5. Specifically, the temperatures of martensite (austenite) start Ms (As) and finish Ms (Af) temperatures determined using DSC (see Fig. 7) are in rough agreement with those that can be evaluated from the magnetization measurements shown in Fig. 6 for sample 5. In contrast, there is no sign of such hysteretic behavior for the other Fe$_2$MnGa samples (not shown).

As can be seen in Fig. 8, at the still higher magnetic field of 50 kOe, there is a total disappearance of the AFM (or FI) order in the FCC phase. Hence, the $M(T)$ plot for sample 2 reveals a typical dependence of the FM ordering with a rather high Curie temperature of $T_C = 800$ K according to the $\chi(T)$ data shown in Fig. 5. On the other hand, the alloys containing mainly the BCC phase (samples 4 and 6) with metamagnetic behavior show significant magnetization much above their Curie temperatures due to the suppression of spin fluctuations by a high magnetic field. In contrast, the $M(T)$ dependence for sample 5 taken at $H = 50$ kOe significantly differs and exhibits the hysteretic behavior characteristic of a martensitic transformation. This is further confirmed by a shift of 20 K for the forward and reverse martensitic transformation temperatures due to a magnetic field in comparison to those determined by DSC (see Fig. 7). In accordance with the magnetization measurements carried out at low magnetic fields (see Fig. 6), the magnetization at $H = 50$ kOe is 20 emu/g at $T > 300$ K, i.e., much lower than $M$ for the alloys containing the BCC or FCC phases.

### Transport properties

Table I shows that the RT values of the resistivity $\rho$ are rather high, perhaps due to both a high resistivity of HA[5] and additional defects caused by the sample preparation. Spark-erosion is known to cause significant distortions of the sample surface to a depth of about 150 $\mu$m.[20] Thus, a distorted portion of a specimen with cross section $1 \times 1$ mm$^2$ occupies about 40% of its volume. Therefore, the values of the resistivity Fe$_2$MnGa alloys should be treated with caution. Nevertheless, in all the investigated Fe$_2$MnGa alloys, the high resistivity correlates with that found in the literature.[1,5]

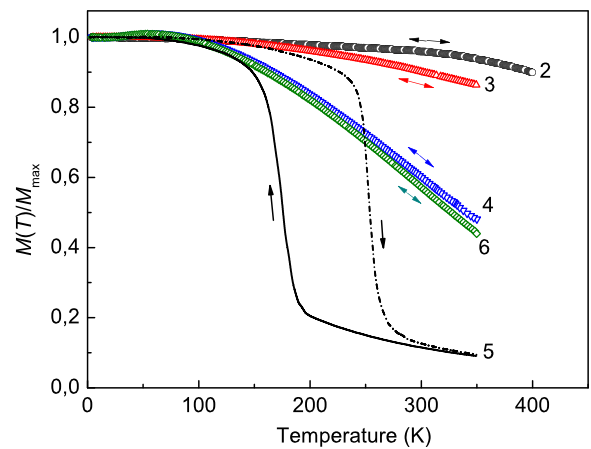

FIG. 8. Temperature dependencies of normalized magnetization of bulk Fe$_2$MnGa alloys taken at $H = 50$ kOe. Numbers near the $M(T)/M_{max}$ curves indicate alloy sample. The $M(T)/M_{max}$ plots for all the samples (except sample 5) practically coincide for the cooling and warming scans, therefore only the warming scans are presented.

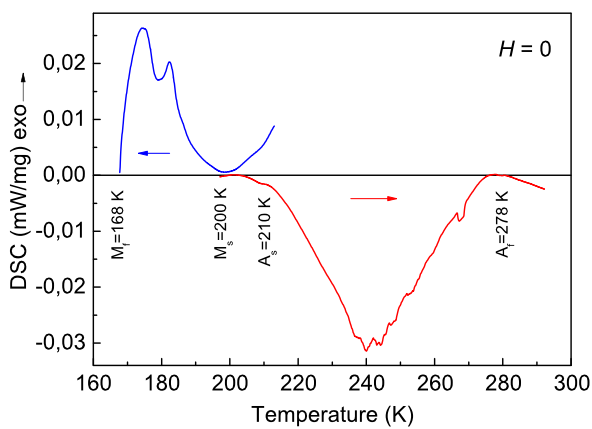

FIG. 7. DSC curves of bulk Fe$_2$MnGa alloy, sample 5. Ms (As) and Mf (Af) are the start and finish temperatures of martensite M (austenite A), respectively.

As shown in Fig. 9, the amounts of each phase (FCC, BCC) in a Fe$_2$MnGa alloy significantly affect the temperature behavior of the resistivity. Sample 1, containing mostly the FCC phase, shows typically metallic behavior: $\rho$ increases nearly linearly with temperature in a temperature range of 80–550 K. An increase in the amount of the BCC phase (samples 2 and 3) results in saturation of the resistivity at a high temperature range. The visible upturn in the $\rho(T)$ for sample 4 takes place at $T \approx 400$ K, where the metallic behavior of the resistivity with a positive temperature coefficient of resistivity (TCR) gives way to a semiconductor-like one with a negative TCR. The temperature dependence of the resistivity for a Fe$_2$MnGa alloy containing mainly the BCC phase (sample 6) exhibits a clear "cusp" at which the TCR changes its sign. This temperature is close to its Curie temperature. A similarly striking change in the character of electron transport as the system undergoes the transition from ferromagnetic to paramagnetic has been found in Fe$_2$MnSi and Co$_2$TiSn HAs.[21–23]

Unlike the apparent similarity with other samples (e.g., samples 4 and 6) both in structure and the phase amount, the transport properties of sample 5 are different. Thus, the resistivity value of sample 5 exhibits (on warming) a rapid and significant increase of about 30% at $T \approx 250$ K, where the TCR changes its sign. Furthermore, only the Fe$_{46}$Mn$_{24}$Ga$_{30}$ alloy (sample 5) has a significant temperature hysteresis of resistivity on heating and cooling (see the inset in Fig. 9). Comparing the obtained experimental $\rho(T)$ dependence of sample 5 with the results in the literature, the nature of the peculiarity at $T \approx 250$ K can be definitely ascribed to a reverse martensitic transformation.[1,5]

## DISCUSSION

The structural characterization of these Fe$_2$MnGa alloys (see Fig. 3) and the first-principles calculations (see Table I) suggest that a relatively small deviation from the stoichiometric composition in Fe$_{50}$Mn$_{25}$Ga$_{25}$ with the FCC structure lead to the emergence of the BCC phase. The BCC phase is ferromagnetic at $T < 220$ K. In contrast, the FCC phase is ferromagnetic at high temperatures ($T_C \approx 730$ K) and at $T < 300$ K becomes ferri- or anti-ferromagnetic. However, this ferri-/anti-ferromagnetic ordering is unstable, and at high magnetic fields of 50 kOe, it experiences a metamagnetic transformation to a ferromagnetic ordering. Additionally, the structural instability of the Fe$_2$MnGa alloys results in a tetragonal distortion of the BCC phase: it becomes energetically more stable in some circumstances due to a martensitic transformation. The martensitic transformation in sample 5 is confirmed by the magnetic (Figs. 6 and 8), DSC (Fig. 7), and partially by the transport (Fig. 10) measurements. A more precise analysis of the XRD data for sample 5 suggests that the occurrence of a tetragonal phase is possible even though the austenite finish temperature Af is at 278 K, i.e., slightly below the RT.

A closer review of the XRD spectra for samples 4, 5, and 6 shows that the diffraction peak of sample 5 located at $2\Theta = 48.2°$ is slightly shifted to the small-angle region in comparison with the (200) reflection for the FCC phase for the other samples. According to Zhu et al.[1] and Omori et al.,[3] the martensite phase has noticeably different lattice parameters ($a = b = 0.5328$ nm, $c = 0.7113$ nm for Fe$_{50.0}$Mn$_{22.5}$Ga$_{27.5}$ alloy or $a = b = 0.5368$ nm, $c = 0.7081$ nm for Fe$_{44}$Mn$_{28}$Ga$_{28}$ alloy) from that of the parent austenite BCC phase ($a = 0.5826$ nm or $a = 0.5864$ nm). The modeling of the XRD spectrum for tetragonal phase by using the experimental data obtained by Omori et al.[3] reveals that the experimental diffraction peaks located at $2\Theta = 42.2$ and $48.2°$ can be nicely fitted with the (202) and (220) reflections in the model XRD spectrum. The presence of the (004), (400), (224), and (422) reflections related to the tetragonal phase can also be further confirmed by comparison of the experimental XRD spectrum with the simulated one. Other intense peaks in the experimental spectrum for sample 5 (e.g., the peaks at $2\Theta = 43.8, 62.8,$ and $80.1°$) are definitely formed by the (220), (400), and (422) reflections of the BCC phase. However, the (111) and (200) reflections from the FCC phase may also contribute to the resulting intensity of the reflections at $2\Theta = 42.2$ and $48.2°$. Therefore, our analysis of the RT XRD pattern of sample 5 suggests that it contains some traces of the tetragonal phase in the BCC matrix. The amount of the tetragonal phase at RT in the sample 5 evaluated using experimental XRD data is close to 20%.

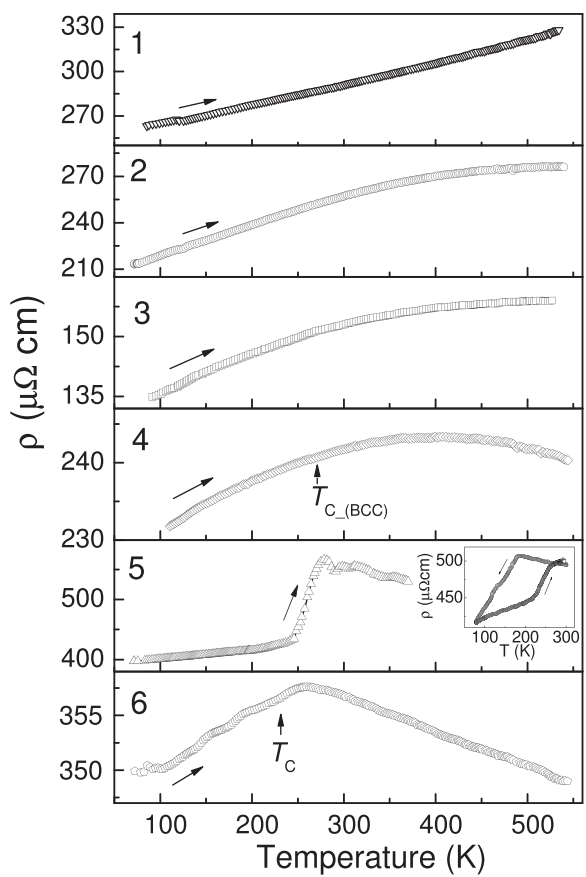

FIG. 9. Resistivity vs. temperature of the Fe$_2$MnGa alloys. Numbers on the panels indicate sample number. Inset in the plot for sample 5 shows a hysteretic behavior of $\rho(T)$ for another sample with the same composition as sample 5.

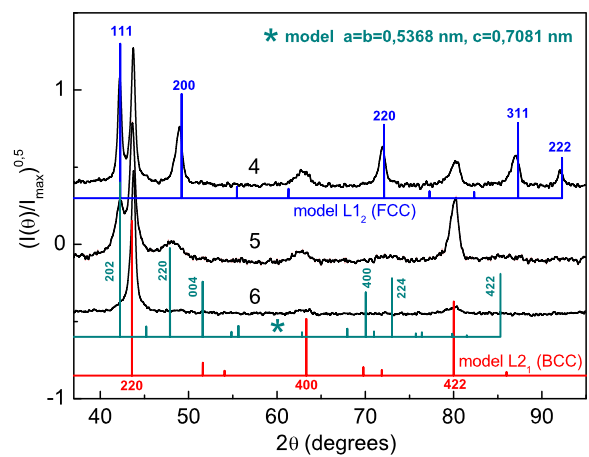

FIG. 10. Experimental XRD spectra of mainly BCC-phase containing Fe$_2$MnGa alloys together with those for L1$_2$, L2$_1$, and tetragonally distorted L2$_1$ lattices.

The question now arises as to how the martensitic transformation happens in $Fe_2MnGa$ alloys for a valence electron concentration $e/a$ below 6.4 (or 6.3 according to our measurements). The complex dependence of the presence of a martensitic transformation on the composition is often approximated by $(e/a)$.[14] However, the universality of this approach is not clear.[14,15] Nevertheless, it is worth noting that, among $Fe_2MnGa$ alloys, a martensitic transformation has been observed earlier exclusively for alloys containing the BCC phase.[1–5] Stoichiometric $Fe_2MnGa$ alloy has $e/a = 6.5$ electrons/atom. It is seen (see Table I) that an increase in the BCC phase content in $Fe_2MnGa$ alloys is accompanied by a decrease in $e/a$. The only $Fe_2MnGa$ alloy with the martensitic transformation has the lowest $e/a = 6.27$. On the other hand, a martensitic transformation for $Fe_2MnGa$ has been previously observed for $6.27 < e/a < 6.40$.[1–5]

It is supposed that the martensitic transformation in Ni-based HA is due to a structural instability driven by their peculiar electron band structure, specifically, due to the formation of an intense $N(E)$ peak at $E_F$.[14,15] For example, in $Ni_2MnGa$ with a BCC structure, the Fermi level $E_F$ lies in a valley of antibonding minority-spin $3d$ states, which leads to a structural instability.[14] This instability can be lifted by transforming to a tetragonal structure because of a gap formed in the minority-spin states around $E_F$. A closer look at the density of states shown in Fig. 2 for $Fe_2MnGa$ with the $L2_1$ ($e/a = 6.5$) and the tetragonal structure (FM or FI) reveals a similar behavior for majority-spin DOS. Assuming that $e/a$ is now 6.22, the Fermi level would be 0.3 eV lower than that for $e/a = 6.5$ and would lie in the DOS peak— clearly an unstable state. Therefore, similarly to what is observed in $Ni_2MnGa$, the instability would be lifted by transforming to a tetragonal structure with the pseudo-gap at $E_F$ depicted by a dashed line in Fig. 2. Such a scenario would explain why sample 5 with $e/a = 6.27$ exhibits a martensitic transformation.

Bearing in mind that our samples contain both BCC and FCC phases and reveal metamagnetism at low temperatures (for the FCC phase), it is shown in Fig. 11 that there is some correlation between $e/a$ and the magnetic moment per formula unit at RT. In Fig. 11, the black circles depict the experimental values of the magnetic moment. The gray circles depict values taken from the literature.[1,3,5,8,11] The dashed red line is a guide to the eye and shows, for low values of $e/a$, the magnetization between $0\,\mu_B$/f.u. of the paramagnetic BCC phase and $2\,\mu_B$/f.u. for the $L2_1$ (FI) structure. For the highest values of $e/a$, the magnetization is lower than $6\,\mu_B$/f.u. for the FCC $L1_2$ phase. It is seen that the magnetization seems to show a jump at $e/a$ of 6.4, but the other processing parameters also play a role,[12] so that samples 4 and 6 do not exhibit any martensitic transformation and have the notably higher magnetization of $1.5\,\mu_B$/f.u. at RT. The magnetic properties of sample 5 differ from those of the other investigated $Fe_2MnGa$ alloys not only in a significant hysteresis on warming and cooling but also in having the lowest magnetization value of $0.5\,\mu_B$/f.u. at $T > 300\,K$ obtained either at low or high magnetic fields (see Figs. 6 and 8). In contrast, in the martensite phase, the magnetization is $3.8\,\mu_B$/f.u. in agreement with the earlier experiments.[2] Such a strange behavior, i.e., transformation from a nearly paramagnetic parent BCC phase to a strongly ferromagnetic tetragonal martensite phase, is the opposite to that of the most known Ni-Mn-based shape memory alloys.[10,24] Therefore, in contrast to a fragile magnetism of the BCC phase in $Fe_2MnGa$ alloys, the magnetism of the tetragonal phase is very stable, in agreement with first-principles calculations (see Table II).

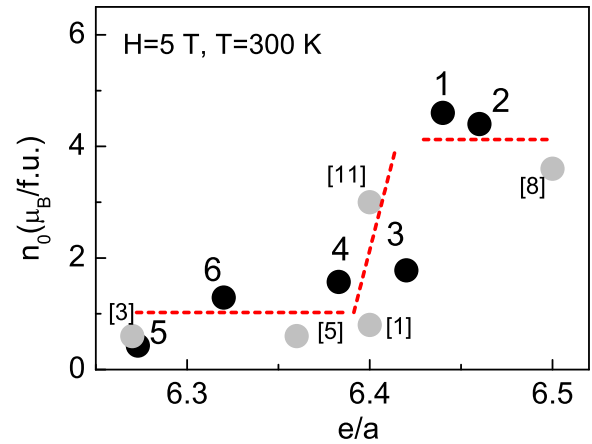

FIG. 11. Magnetization (in $\mu_B$ per formula unit) vs. the valence electron concentration $e/a$ of the investigated $Fe_2MnGa$ alloys (black circles and numbers) and the results taken from the literature (gray circles with reference numbers). The red dashed line is a guide to the eye.

It has been shown that a change in FCC to BCC amounts in $Fe_2MnGa$ alloys causes the gradual changes in $\rho(T)$ dependencies from the metallic-like to the semiconductor-like behavior (see Fig. 10). The martensitic transformation observed in sample 5 supplements this tendency by a rapid increase in resistivity in a narrow temperature range (see Fig. 10). The resistivity described within a simple relaxation time $\tau$ approximation in a nearly free electron model (see, for example, Ref. 25) reads as

$$\rho = \frac{m^*}{e^2 \tau n_{\text{eff}}} = \frac{3}{e^2 \tau v^2} \frac{1}{N(E_F)}, \quad (1)$$

where $m^*$ and $n_{\text{eff}}$ are the effective mass and the effective density of conduction electrons per unit volume, respectively; $v$ and $N(E_F)$ are the velocity and the DOS at the Fermi energy. According to results of the first-principles calculations (see Table II and Fig. 2), the value of $N(E_F)$ for the tetragonal phase is nearly five times higher than that for the $L2_1$ phase. Thus, the experimentally observed decrease in resistivity upon transformation from austenite to martensite can be explained by a significant increase in $N(E_F)$ for the tetragonal phase. The values of $N(E_F)$ for the L1$_2$ phase with FM or FI types of magnetic order are more than three times larger than that for the L2$_1$ phase. This also correlates with the experimentally observed dependencies of $\rho(T)$ of those Fe$_2$MnGa alloys that exhibit a martensitic transformation.

Several mechanisms of charge carrier scattering have been suggested to explain the temperature dependence of the resistivity of metals. Electron–phonon, electron–electron, and electron–magnon (spin-disorder) scattering make major contributions to the resulting temperature dependence of ferromagnetic metals with positive TCR. Thus, for example, electron–phonon based scattering ($\sim T$ for $T \gg \Theta_D$) can be described by the Bloch–Grüneisen function and its modifications.[26] Electron–electron scattering is proportional to $\sim T^2$.[27] Electron–magnon (spin-disorder) scattering mechanisms are proportional to $\sim T^\alpha$, where $\frac{3}{2} < \alpha < T^{\frac{9}{2}}$.[28] Electron–magnon (spin-disorder) scattering usually reaches its maximum near the Curie temperature; above $T_C$, the spin-disorder mechanism is independent of $T$. Therefore, for some ferromagnetic metals and alloys (including ferromagnetic HA[23]), a distinct change in the slope of the $\rho(T)$ dependence can be expected above $T_C$.[29–31]

The Ioffe–Regel criterion predicts the conditions when the resistivity is saturated with temperature: when the mean free path $l$ of the electrons is of the interatomic distance $d$

$$\rho^{\text{I-R}} = 55 \times \frac{d}{r_B} (\mu\Omega\text{cm}), \quad (2)$$

where $d$ is the nearest interatomic distance and $r_B = 0.0529$ nm is the Bohr radius.[32] For the FCC and BCC phases of Fe$_2$MnGa alloys, $d$ is 0.262 and 0.254 nm, respectively. Thus, $\rho^{\text{I-R}}$ for Fe$_2$MnGa alloys should be 272 ÷ 264 $\mu\Omega$ cm. The experimentally determined values of the alloy resistivity are close to these parameters. However, unlike the expected change of the slope in $\rho(T)$ at $T_C$ or even its saturation above $T_C$, for the high-resistive Fe$_2$MnGa HA (samples 4, 6), a negative slope of $\rho(T)$ at high temperatures is also observed in Fig. 9. A similar behavior of $\rho(T)$ at $T > T_C$ has also been observed for the other HAs.[21–23,31]

There are several reasons for the appearance of the negative TCR in metals. The negative TCR of some transition metal alloys is usually attributed to the strong atomic scattering's leading to a weak localization or variable-range-hopping.[33,34] Mooij[35] has suggested an empirical rule for such alloys: the TCR is negative if the resistivity of the alloy exceeds $\rho_{\text{alloy}} > 150\ \mu\Omega$ cm. The temperature dependence of the resistivity of sample 6 with the BCC structure shows a nearly linear decrease with temperature above $T_C$ (Fig. 9). The quantum interference effects such as weak localization and electron–electron interactions give rise to a rather high value of $\rho$ and a negative TCR in disordered systems.[33] Furthermore, the variable-range-hopping conductivity[34] can also lead to a negative TCR. In the variable-range-hopping mechanism proposed by Mott,[34] electrons preferentially hop to the localized states that are close in energy but not necessarily close spatially, and the conductivity is

$$\frac{1}{\rho(T)} = \sigma(T) = \sigma_0 \exp\left[-\left(\frac{T_0}{T}\right)^p\right], \quad (3)$$

where $p = 1/4$. The weak localization mechanism shows that the conductivity depends on the temperature

$$\Delta\sigma = \alpha(2e^2/\pi^2\hbar)\ln T, \quad (4)$$

where the coefficient $\alpha$ is negative.[33]

As shown in Fig. 12, the temperature dependence of the conductivity of sample 6, containing mostly the BCC phase, changes at $T_C$ of 240 K (i.e., $\ln(240) = 5.48$) in accordance with the weak localization mechanism. However, it is worth noting that the variable-range-hopping mechanism also gives a reasonable approximation to the experimental data (not shown).

Among the investigated Fe$_2$MnGa alloys, sample 4 also contains a significant amount of BCC phase (about 60%) with a Curie temperature of 250 K. However, no peculiarity near $T_C$ can be seen. This is most probably due to a bridging by a lower resistivity FCC phase. For Fe$_2$MnGa alloys rich in the FCC phase (samples 1–3), the metamagnetic transformation from AFM (or FI) to FM takes place at $T \approx 300 \div 350$ K (on warming, see Fig. 3). However, there is no peculiarity in $\rho(T)$ in this range of temperatures. Thus, for these high-resistive Fe$_2$MnGa alloys, spin-disorder scattering makes a negligible contribution in comparison with the other scattering mechanisms.

## SUMMARY

The experimental results for Fe$_2$MnGa alloys shown in this paper supplement those previously obtained, as resulting from their metastable behavior. There are some important points aiming at an explanation of the variety of magnetic and electronic transport properties of Fe$_2$MnGa alloys.

(1) The first-principles calculations for the stoichiometric Fe$_2$MnGa alloy with different types of atomic and magnetic orders show that the L2$_1$ structures with the ferrimagnetic ordering as well as the L1$_2$ structure with the ferromagnetic ordering are the most stable structures. In contrast, the tetragonally distorted L2$_1$ phase with any type of magnetic order is less probable for the stoichiometric Fe$_2$MnGa alloy.

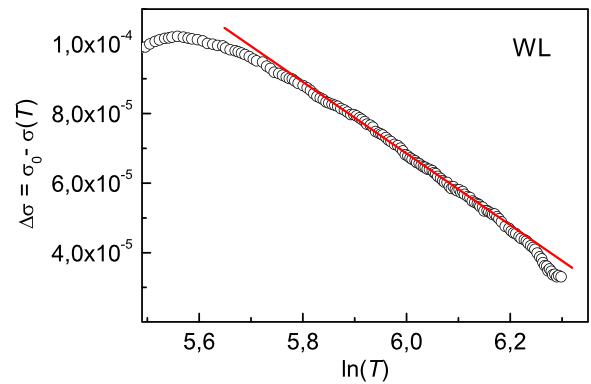

FIG. 12. Change of the electrical conductivity vs. $\ln(T)$ of a Fe$_2$MnGa alloy containing mostly the BCC phase (sample 6) for $T > T_C$. Red continuous line shows a fit according to Eq. (4) for a weak localization mechanism.

(2) Several single or two-phase off-stoichiometric Fe$_2$MnGa alloys have been investigated with $6.25 < e/a < 6.5$.

(3) It has been shown that Fe$_2$MnGa alloys with nearly pure BCC or FCC phases have distinct magnetic properties, with FM–PM and AFM–FM–PM transformations, respectively. The Fe$_2$MnGa alloys with mixed BCC and FCC phases have the features typical for both BCC and FCC phases.

(4) The transport properties of the Fe$_2$MnGa alloys containing FCC and BCC phases correlate with the results of first-principles calculations of these phases. The negative TCR of a Fe$_2$MnGa alloy containing a pure BCC phase can be explained by a weak-localization mechanism.

(5) We have found that the Fe$_{46}$Mn$_{24}$Ga$_{30}$ alloy with the lowest value of valence electron concentration ($e/a = 6.27$) contains predominantly the BCC phase and has a martensitic transformation with $M_s = 168$ K and $A_s = 225$ K for $H = 0$ as well as $M_s = 193$ K and $A_s = 237$ K for $H = 50$ kOe. The transformation is accompanied with significant changes in its magnetic and transport properties. The martensite phase is in a FM state, while the parent phase is in a PM state.

## ACKNOWLEDGMENTS

This work has been supported by the project "Marie Skłodowska-Curie Research and Innovation Staff Exchange (RISE)" Contract No. 644348 with the European Commission, as part of the Horizon2020 Programme.

The authors are deeply appreciative of the help of A. V. Terukov and M. S. Nizameev in the XRD and DSC measurements.